\documentclass[twocolumn,prl,10pt,amsmath,amssymb,nofootinbib,showpacs,superscriptaddress,floatfix]{revtex4-1}

\DeclareFontFamily{U}{rcjhbltx}{}
\DeclareFontShape{U}{rcjhbltx}{m}{n}{<->rcjhbltx}{}
\DeclareSymbolFont{hebrewletters}{U}{rcjhbltx}{m}{n}

\newcommand{\rs}{\rm\scriptscriptstyle}

\usepackage{graphicx}
\usepackage{color}
\usepackage[usenames,dvipsnames]{xcolor}
\usepackage[colorlinks=true,linkcolor=Red,citecolor=Green,linktoc=page]{hyperref}
\usepackage{multirow}
\usepackage{float}
\usepackage{flushend}
\usepackage{balance}
\usepackage[varg]{txfonts}
\usepackage{ulem}
\usepackage{fancyhdr}

\DeclareMathSymbol{\lamed}{\mathord}{hebrewletters}{108}

\begin{document}
\title{Topological Nature of High Temperature Superconductivity}
\author{	M.\,C.\,Diamantini}
\affiliation{NiPS Laboratory, INFN and Dipartimento di Fisica e Geologia, University of Perugia, via A. Pascoli, I-06100 Perugia, Italy}
\author{C.\,A.\,Trugenberger}
\affiliation{SwissScientific Technologies SA, rue du Rhone 59, CH-1204 Geneva, Switzerland}
\author{V.\,M.\,Vinokur}
\affiliation{Terra Quantum AG, St. Gallerstrasse 16A, CH-9400 Rorschach, Switzerland}
\affiliation{University of Twente, Faculty of Science and Technology and MESA+ Institute of Nanotechnology, \\7500 AE Enschede, The Netherlands}

\begin{abstract}
The key to unraveling the nature of high-temperature superconductivity (HTS) lies in resolving the enigma of the pseudogap state.  
The pseudogap state in the underdoped region is a distinct thermodynamic phase characterized by nematicity, temperature-quadratic resistive behavior, and magnetoelectric effects. Till present, a general description of the observed universal features of the pseudogap phase and their connection with HTS was lacking. The proposed work constructs a unifying effective field theory capturing all universal characteristics of HTS materials and explaining the observed phase diagram. The pseudogap state is established to be a phase where a charged magnetic monopole condensate confines Cooper pairs to form an oblique version of a superinsulator. The HTS phase diagram is dominated by a tricritical point (TCP) at which the first order transition between a fundamental Cooper pair condensate and a charged magnetic monopole condensate merges with the continuous superconductor-normal metal and superconductor-pseudogap state phase transitions. The universality of the HTS phase diagram reflects a unique topological mechanism of competition between the magnetic monopole condensate, inherent to antiferromagnetic-order-induced Mott insulators and the Cooper pair condensate. The obtained results establish the topological nature of the HTS and provide a platform for devising materials with the enhanced superconducting transition temperature.
\end{abstract}

\maketitle

\section{Introduction}

The origin and underlying mechanism of high-temperature superconductivity (HTS) has been remaining a subject of heated debate for more than three decades after its discovery. Even the metallic state above the superconducting transition temperature $T_{\mathrm c}$ seem to differ strikingly from metals in which conventional superconductors transform\,\cite{Barisic2013,Zaanen2015,Matsuda2017,Bozovic2019,Taillefer2019}. In the underdoped regime of HTSs the pseudogap state (PG) opens up at the temperature $T^{*}$
much above $T_{\mathrm c}$, see\,\cite{Taillefer2019}
and references therein. The  PG state exhibits striking properties including nematicity\,\cite{Matsuda2017}, temperature-quadratic resistive behavior, magnetoelectric effects\,\cite{Barisic2013}, and carries a signature of pairing well above $T_{\mathrm c}$\,\cite{Bozovic2019}. These exotic behaviors suggest that understanding the nature of HTS requires gaining a fundamental insight into the nature of the PG state.
\bigskip

\begin{figure}[t!]
	\includegraphics[width=9cm]{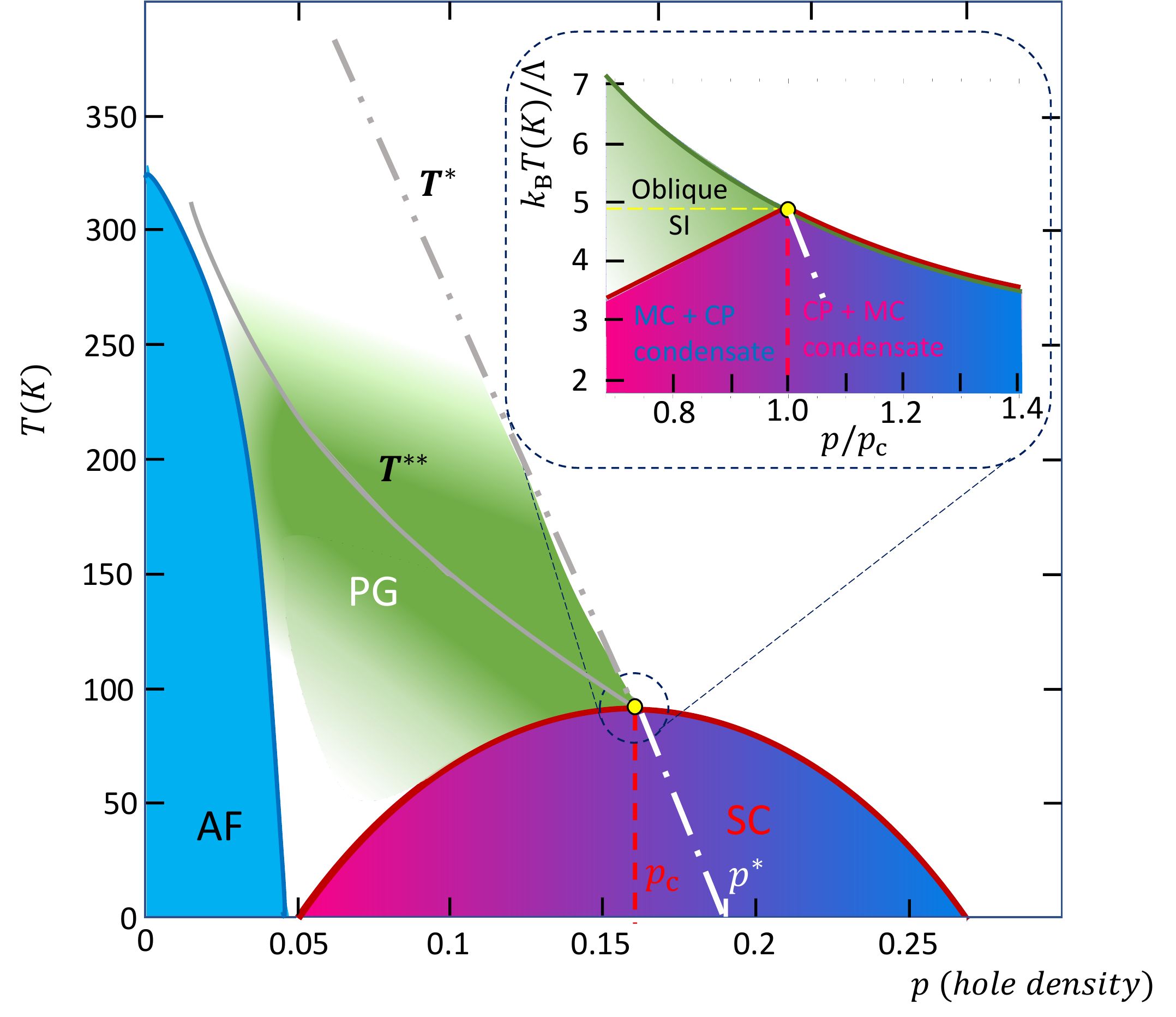}
	\vspace{-0.2cm}
	\caption{\textbf{A sketch of the temperature-doping phase diagram of HTS.} The phase diagram presents different phases, including antiferromagnetism (blue), superconductivity (purple to dark blue), and pseudogap (green) domains. The dashed gray line at $T^*$ marks the onset of the pseudogap phase detected by deviation from the linear $R_{\rs\square}\propto T$ resistance dependence. The gray dashed-point line, $T^{**}$, indicates the emergence of square $R_{\rs\square}\propto T^2$ resistance dependence, nematicity, and magnetoelectric effects. The $p_{\mathrm c}$ marks the tricritical point at the dome maximum, and $p^{\ast}$, where $T^*$ hits the x-axis, is the position of the putative quantum Mott-insulator-superconductor phase transition at $T=0$. \textbf{Inset:} The calculated phase transition lines in the vicinity of the tricritical point. 
	}
	\label{Fig1}
\end{figure}

The behavior of HTS materials is graphically summarized in the phase diagram in Fig.\,\ref{Fig1}. At very low doping, strong electron interactions in HTS, responsible for strong Cooper pairing\,\cite{Taillefer2019}, localize the antiferromegnetic state into a Mott insulator\,\cite{Barisic2013,Proust2016,Davis2019}.  Further doping, i.e. adding $p$ holes per Cu atom, turns the electrons itinerant. Thus, the superconductivity dome forms in the $p\approx 0.05\div 0.3$ interval, followed by a Fermi liquid at higher doping, see phase diagram in Fig.\,\ref{Fig1}. The elusive PG state emerges in the underdoped region. Above the temperature $T^*$, associated with the onset of the PG, a metallic sheet resistance linear in temperature, $R_{\rs\square}\propto T$, is observed. At the same time, below temperature, $T^{**}<T^*$, a distinct switch to a quadratic  $R_{\rs\square}\propto T^2$ dependence occurs, holding till superconducting fluctuations set in, see\,\cite{Barisic2013,Proust2016} and references therein. The full switching on of the PG state  at $T$$=$$T^{**}$ is accompanied by a nematic phase transition and the emergence of the magnetoelectric (Kerr) effect, see\,\cite{Kivelson1998,Kivelson2014,Matsuda2017,Davis2019} and references therein, evidencing that the PG state is a distinct thermodynamic phase. Recent shot noise measurements deep in the PG region detected pairing of the charge carriers\,\cite{Bozovic2019}. Finally, the evolution from the Mott insulator at $p=0$ to superconductivity upon increasing $p$ suggests a quantum superconductor-insulator transition (SIT)\,\cite{critical} at the putative transition point $p^{\ast}$ where the $T^*(p)$ line continued into the dome hits the $p$-axis, see Fig.\,\ref{Fig1}. 
\bigskip

Building on the idea that it is quantum criticality associated with the SIT that dominates the physics of HTS and neighboring phases, we adopt the machinery of the topological gauge theory of the SIT\,\cite{dst, dtv1} for the description of the HTS phase diagram. Deriving from the established Fermi-liquid-like $R_{\rs\square}\propto T^2$ behavior of the resistance in the PG regime, combined with the fact that the dominant charge carriers in the pseudogap state carry the charge $2e$, we demonstrate that the PG regime represents a novel state, the \textit{oblique superinsulator}, in which the electric conductance is mediated by symmetry-protected surface fermionic modes. This enables us to construct a theory that encompasses all the observed features of the PG and HTS phases within a unified universal description.
We show that the key ingredient of the HTS mechanism is the formation of a condensate of dyons, particles carrying both electric and magnetic charge, which appear as topologically-charged magnetic monopoles at the end of open vortices playing the role of Dirac strings\,\cite{dtv3, dtv4}. The superconducting dome of HTS appears as a coexistence phase of Cooper pair and dyon condensates. The PG state above $T_c$ in the underdoped region is a pure dyon condensate characterized by magnetoelectric and nematic effects, in a perfect agreement with the experiment. In this state there are symmetry-protected fermionic surface dyons of charge $2e$ which live on a bubble-like texture and, as a result, cause the observed quadratic in $T$ resistance. The high transition temperature of the HTS is explained by the topological obstruction of splitting Cooper pairs into single electrons in presence of magnetic monopoles of half the fundamental magnetic charge, i.e. by the Dirac quantization condition. 

\section{The model and free energy of the dyonic state}~~

We start constructing our model with stating its two key ingredients, the relevance of the SIT for the HTS mechanism and the established Fermi liquid-like behaviour of the charge carriers in the PG state\,\cite{Barisic2013}, which implies that they are symmetry-protected surface states. Three possible states involved in the SIT are superconductors, bosonic topological insulators (BTI)\,\cite{bm} and superinsulators\,\cite{dst, vinokur, dtv1}. Superconductors do not carry surface states. In three dimensions, the BTI, in which neither time-reversal symmetry breaking nor topological order occur, also cannot provide the necessary fermionic surface degrees of freedom\,\cite{senthil}. The PG state should thus be associated with superinsulators. Recalling that the pseudogap state exhibits also the magnetoelectric effect\,\cite{Barisic2013}, we conclude that the Lagrangian describing the electromagnetic response of the system should include the so-called $\theta$-term, ${\cal L}_{\theta}$. As we now demonstrate,  the $\theta$-term\,\cite{axion} transforms the standard superinsulator into a new state, which we call oblique superinsulator, and which does indeed host symmetry-protected fermionic surface dyons of charge $2e$. 
\bigskip

The $\theta$-term is given by the expression
\begin{equation}
	{\cal L}_{\theta} = {\theta \over 32 \pi^2} F_{\mu\nu} \epsilon^{\mu \nu \alpha \beta} F_{\alpha\beta} = {\theta \over 4\pi^2} {\bf E} \cdot {\bf B} \,,
	\label{theta}
\end{equation}
where $F_{\mu\nu}=\partial_{\mu} A_{\nu}-\partial_{\nu}A_{\mu}$ is the electromagnetic tensor, $A_{\nu}=(\varphi,-{\bf A})$ is the gauge potential, $(\theta/4 \pi^2)\delta_{ij}$ represents the pseudoscalar part of the magnetoelectric coupling, or polarizability $\alpha_{ij}\equiv[\partial M_{j}/\partial E_{i}]_{\mathbf{B}=0 }\equiv[{\partial P_i/\partial B_j}]_{\mathbf{E}=0}$, ${\bf E}=-\nabla\varphi-\partial_t {\bf A}$ and ${\bf B}=\nabla\times{\bf A}$ are the electric and magnetic fields, respectively, ${\bf P}$ and ${\bf M}$ are the respective polarization and magnetization. Here we use natural units $c$$=$$1$, with $c$ being the light velocity in the material, $\hbar$$=$$1$, $\varepsilon_0$$=$$1$, Greek letters denote space-time indices and a sum is implied over repeating indices. Since the $\theta$-term can be presented as a full derivative, $\propto \partial_{\mu}(\epsilon^{\mu\nu\alpha\beta}A_{\nu}\partial_{\alpha}A_{\beta})$, it is a pure surface term on bounded spaces. On a compact Euclidean torus ${\bf T}^4$, the integral of the $\theta$-term is a topological invariant contributing a factor ${\rm exp}( i n\theta) $ to the partition function, hence $\theta$ is 2$\pi$-periodic. The $\theta $-term describes the topological coupling of a vector and a pseudovector field, therefore, it is odd under parity and time-reversal symmetries. The values of $\theta$ compatible with the time-reversal symmetry are thus $\theta = 0\  ({\rm mod}\ 2\pi)$ and $\theta = \pi \ ({\rm mod}\ 2\pi)$. 
\bigskip

The electromagnetic response of superinsulators is described\,\cite{dtv1} by the compact QED\,\cite{polyakov} in the phase where magnetic monopoles\,\cite{olive} form a Bose condensate\,\cite{dtv3, dtv4}. The inclusion of the $\theta$-term in the compact QED theory results in the Euclidean lattice gauge model\,\cite{cardy}
\begin{eqnarray}
	&&Z = \sum_{ \{ n_{\mu} \} , \{ m_{\mu} \} } \int {\cal D} A_{\mu} {\rm e}^{-S} \ ,
	\nonumber \\
	&&S = \sum_x {1\over 4f^2} \left( F_{\mu \nu} -2\pi S_{\mu \nu} \right)^2 + i  A_{\mu} \left( n_{\mu} + {\theta \over 2\pi} m_{\mu} \right)  \ ,
	\label{compactqed}
\end{eqnarray}
where the integers $n_{\mu}$ and $m_{\mu} = (1/2) K_{\mu \alpha \beta} S_{\alpha \beta}$ are the conserved charge and magnetic monopole currents, respectively, $K_{\mu \alpha \beta}$ is the lattice BF operator\,\cite{dst} (see Methods), and the dimensionless coupling $f$ is an effective strength of the Coulomb interaction in the material. Magnetic monopoles are topological excitations arising in the electromagnetic response due to the compact nature of the electromagnetic gauge potential. Equation\,(\ref{compactqed}) implies that, in the presence of the topological $\theta$-term, magnetic monopoles acquire an electric charge $\theta/2\pi$ and become {\it dyons}. This is known as Witten effect\,\cite{witteneffect}. 
Having in mind Cooper pairs, we let the electric charge unit be $2e$. Dirac quantization $qg/(4\pi\hbar)=n/2$, with $q$ and $g$ being electric and magnetic charges, and an integer $n$\,\cite{olive}, requires the unit of magnetic charge be $\pi/e$. 
\bigskip

Integrating out the gauge field $A_{\mu}$, one obtains the model formulated solely in terms of charge and monopole currents,
\begin{equation}
	Z = \sum_{ \{ n_{\mu} \} , \{ m_{\mu} \} } {\rm e}^{-S} \ ,
\end{equation}
\vspace{-0.3cm}
\begin{eqnarray}
	S = \sum_x f^2 \left( n_{\mu} +{\theta \over 2\pi} m_{\mu} \right)  {1\over -\nabla^2} \left( n_{\mu} 
	+{\theta \over 2\pi} m_{\mu} \right)+ {\pi^2 \over f^2} \sum_x m_{\mu} {1\over -\nabla^2} m_{\mu} \nonumber \\
	+ i \pi \ n_{\mu} {1\over -\nabla^2} K_{\mu\alpha\beta} M_{\alpha \beta} \ ,
	\label{charmon}
\end{eqnarray}
where the integers $M_{\mu \nu}=(1/2) S_{\mu} \epsilon_{\mu \nu \alpha \beta} S_{\alpha \beta}$ are such that $m_{\mu} = \Delta_{\nu} M_{\mu \nu}$ ($S_{\mu}$ is the lattice shift operator, see Methods). The last term represents the topological interaction between charges and monopoles. When the Dirac quantization condition is satisfied, it falls out of the partition function since it is always an integer multiple of $2\pi$, as can be easily seen by representing the conserved charge current as $n_{\mu} = K_{\mu \alpha \beta} X_{\alpha \beta}$, with the gauge conditions $\Delta_{\alpha} X_{\alpha \beta} = \Delta_{\beta} X_{\alpha \beta} = 0$, so that $X_{\alpha \beta}$ contains four degrees of freedom as $n_{\mu}$, and using the relation (\ref{ma}) in Methods. 

\bigskip
Keeping only self-interaction terms, one finds that the action associated with a particle is proportional to the length $N$ of its Euclidean world-line, representing a lattice string\,\cite{kogut}. As usual in a statistical field theory, this Euclidean action is equivalent to the ``energy" of an equivalent statistical model in one more spatial dimension, where the coupling constant plays the role of ``temperature". Since the entropy of a lattice string is also proportional to its length, the ``free energy" of a particle with the electric charge $n$ and magnetic charge $m$ becomes
\begin{equation}
	F= \left[ f^2 G(0) \left( n+{\theta \over 2\pi} m\right)^2 + {\pi^2 \over f^2} G(0) \ m^2 -\mu \right] N \ ,
	\label{freeenergy}
\end{equation}
where $G(0)$ is the value of the lattice Coulomb potential at coinciding points and $\mu \approx {\rm ln 7}$, since at each step the non-backtracking string has to choose among 7 possible directions to continue. if the factor in the brackets is positive, the free energy is minimal for $N=0$. If, instead, it is negative, the free energy is minimized by $N=\infty$. This means that particles with quantum numbers resulting in the positive factor in brackets in Eq.\,(\ref{freeenergy}) are suppressed and exist only as short-living fluctuations, while particles with quantum numbers for which the factor in brackets is negative form Bose condensates. If both Bose condensates are possible, the one with the lowest free energy is stable, but the phases may coexist. The condensation condition 
\begin{equation}
	\eta {f^2\over \pi} \left( n+{\theta \over 2\pi} m \right)^2 + \eta {\pi \over f^2}   m^2 < 1 \ ,
	\label{ellipse}
\end{equation}
where $\eta = \pi G(0)/\mu$, describes the interior of a tilted ellipse with the semiaxes $r_n$$= $$\sqrt{\pi/f^2} \sqrt{1/\eta} $ and $r_m$$=$$\sqrt{f^2/\pi} \sqrt{1/\eta}$ on an integer lattice of electric and magnetic charges. The quantity $1/\sqrt{\eta}$ defines the overall scale of the ellipse, while the coupling $f^2/\pi$ is the ratio between its semiaxes. 
Varying the coupling $f$ and angle $\theta$ leads to complex phase structures consisting of alternating sequences of three possible phases\,\cite{cardy}, a superconducting phase consisting of a pure charge condensate ($m$$=$$0$), a Coulomb phase with no condensate, and an oblique confinement phase, where condensed particles carry both electric and magnetic charges\,\cite{thooft}.

\begin{figure*}[t!]
	\includegraphics[width=16cm]{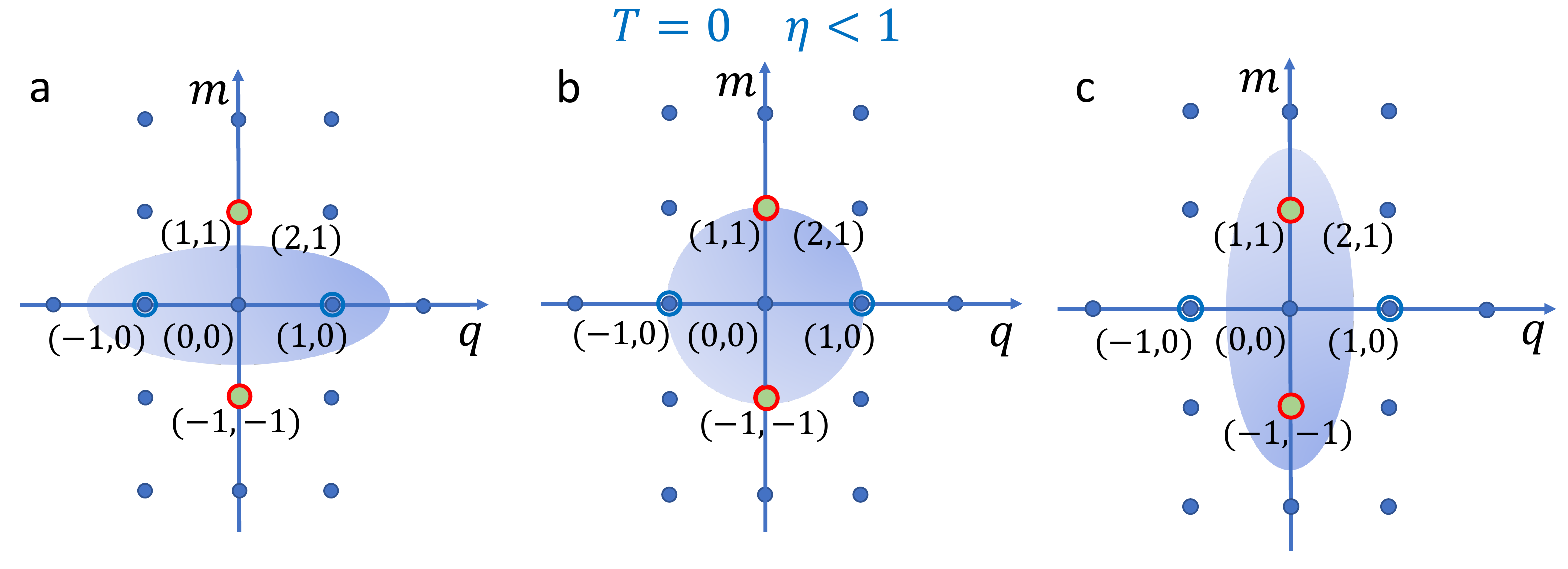}
	\vspace{-0.5cm}
	\caption{\textbf{The ellipse technique for calculating emerging phases at $T=0$ and $\eta<1$}. The $x$-axis shows the integer dyonic charges $q=m+n$, for $\theta=2\pi$, while the $y$-axis lists integer magnetic charges $m$.  
		\textbf{a:} Superconducting state, where only points $n=\pm 1$ (electric charge) and $m=0$ (blue circles) fall inside the ellipse. \textbf{b:} First order phase transition between the superconductor and oblique superinsulator as the ellipse contains both superconducting and dyonic unit charge ($q=\pm 1$, $m=\pm 1) $ integer points (red rim green circles). \textbf{c:} Oblique superinsulator phase.}
	\label{Fig2}
\end{figure*}

\begin{figure}[t!]
	\includegraphics[width=8.5cm]{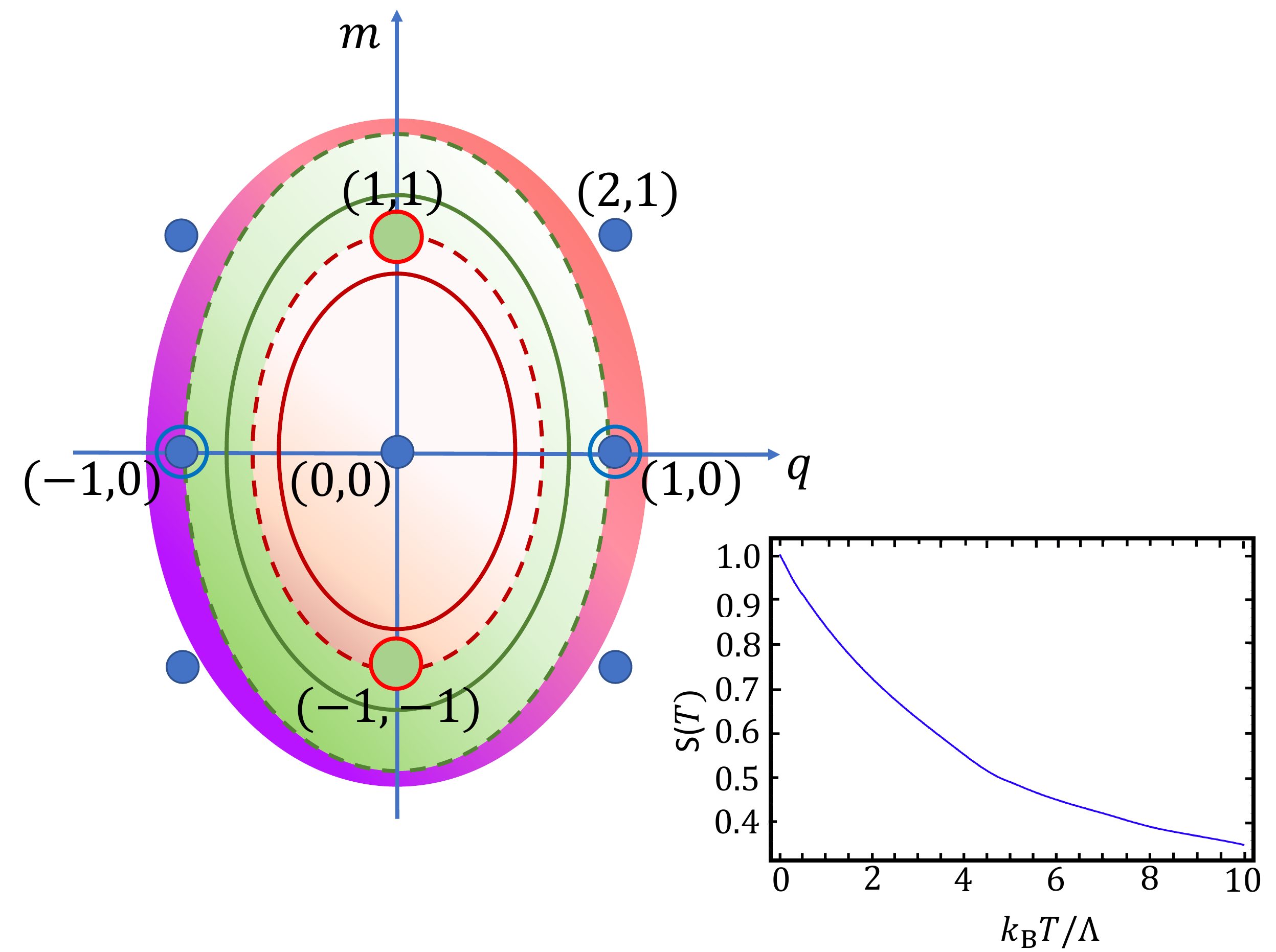}
	\vspace{-0.2cm}
	\caption{\textbf{Temperature evolution and deconfinement at $p$$<$$p_{\mathrm c}$.}
		The purple ellipse corresponds to the coexistence of dyonic and superconducting condensates. The green dashed line corresponds to the temperature at which the transition from the superconducting dome to the pseudogap phase occurs. The solid green line corresponds to temperatures at which the pseudogap phase exists. The dashed fire brick line marks the further transition from the pseudogap to a metallic state and the solid fire brick line coresponds to the metallic state. \textbf{Inset:} The scaling function $S(T)$, see Methods.}
	\label{Fig3}
\end{figure}

\section{The phase diagram and HTS mechanism}
The seminal reference\,\cite{cardy} restricted its analysis to $\theta \in [0, 2\pi [$ because of the periodicity of the charge spectrum under shifts $\theta$$\to$$\theta + 2\pi$. This periodicity is evident in Eqs.\,[\ref{freeenergy},\ref{ellipse}], since the shift $\theta$$\to$$\theta + 2\pi$ can be compensated by the corresponding shift $n$$\to$$n-m$. As a consequence of this restriction, the superconducting and oblique confinement phases can never be adjacent, but are always separated by the Coulomb phases\,\cite{cardy}. Importantly, however, the correct periodicity in $\theta$ is $4\pi$. While the charge of dyonic composites is indeed periodic under shifts $\theta \to \theta +2\pi$, their {\it statistics} is not, if charges are bosons, as has been pointed out in\,\cite{fisher}. The statistics of a composite of $n$ bosonic electric charges and $m$ magnetic monopoles with $\theta=0$ is $(-1)^{nm}$, where $+1$ corresponds to bosons and $-1$ to fermions\,\cite{ds1,ds2}. If we turn on the angle $\theta$, the total electric charge changes to $q=n+m\theta / 2\pi$ but the statistics remains unchanged\,\cite{goldhaber}. If we compensate the shift $\theta \to \theta +2\pi$ by the corresponding transformation $n \to n-m$, to leave the total charge unchanged, we change the statistics of the composite by the factor $(-1)^{m^2}$, which is not necessarily unity. Only the shift $\theta \to \theta + 4\pi$, accompanied by $n \to n-2m$,  yields the same charge {\it and} statistics. Had charges been fermions, then the correct periodicity, indeed, would have been $\theta \to \theta + 2\pi$, but fermions do not Bose condense. This statistical Witten effect\,\cite{fisher} is the same reason why strong bosonic topological insulators are also periodic only under shifts $\theta \to \theta + 4\pi$, contrary to fermionic ones\,\cite{senthil}. 
\bigskip

Since the correct periodicity in the angle $\theta$ is $4\pi$, and not $2\pi$, the case $\theta = 2\pi$ is generically different from that with $\theta  = 0$. For $\theta = 2\pi$ the $(n,m)$ lattice is not tilted, but all $m= {\rm const \ne 0}$ lines are shifted in their $n$-values with respect to the x-axis $m=0$, as shown in Fig.\,\ref{Fig2}. Cranking down the coupling $\pi/f^2$ from high (corresponding to weak Coulomb interactions) to small (corresponding to strong Coulomb interactions) values, first elongates the ellipse along the x-axis, then passes it through a circle, and, finally, elongates it along the y-axis. At $\eta > 1$, the nearly circles forming at the mid-way contain only the origin of the lattice, which corresponds to the Coulomb insulating phase. At $\eta < 1$, instead, the deformation processes from horizontally elongated ellipses containing the lattice point $(n=1, m=0)$ and describing the superconductor, Fig.\,\ref{Fig2}a, through a circle containing {\it both kinds} of lattice points, Fig.\,\ref{Fig2}b, to vertically elongated ellipses containing the lattice point $(n=0, m=1)$ and corresponding to the oblique confinement, Fig.\,\ref{Fig2}c. This describes a first-order transition at which the superconducting order and oblique confinement coexist.  The particles in both these coexisting condensates carry unit charge, but for the `superconducting' particles the charge is fundamental, while `oblique confinement' carriers acquire charge due to the Witten effect. Since $G(0)=0.155$ and $\mu \approx {\rm ln 7}$, we have $\eta$$=$$0.25$$<$$1$. It is exactly this coexistence regime that is realized, without an intervening normal insulating phase. Upon increasing the temperature, the overall scale of the ellipse shrinks, see Fig.\,\ref{Fig3}, by the scale factor $S(T)$ shown in the inset in Fig.\,\ref{Fig3}, see Methods. Thus, lattice points that were within the ellipse at $T=0$ fall out as the temperature grows and the system crosses from the under-the-dome HTS coexistence phase to the pseudogap- and then to a metallic phase as illustrated in Fig.\,\ref{Fig3}. 
\bigskip

As we have already mentioned, the phase harboring the magnetic monopole condensate is a superinsulator\,\cite{dst, vinokur, dtv1, dtv2}. We, therefore, name the material hosting an oblique confinement phase an {\it oblique superinsulator}. 
The numerically computed phase diagram in the vicinity of the tricritical point is shown in the inset in Fig.\,\ref{Fig1}, where we
have identified the coupling constant $\pi/f^2$ with the doping normalized to unity at the $T_{\mathrm c}$ maximum.
The values of $T_{\mathrm c} = {\cal O}(100)$\,K are obtained for the cutoff $\Lambda \approx 2 $\,KeV, i.e. frequencies $\approx 1/2$\,THz which is of order of the plasma frequency in cuprates. The colored regions mark a charged monopole condensate with the magnetic charge $\pi/e$. Accordingly, stable single electrons are forbidden by the Dirac quantization condition, since they would have carried the angular momentum 1/4\,\cite{olive}. In finite systems with a finite density of monopoles this topological obstruction reduces to a finite energy barrier, and single electrons can appear only as short-lived fluctuations with energies above this barrier. Below the superconducting dome (purple), the charged monopole condensate coexists with the fundamental Cooper pair condensate, and the dash-dotted line hitting $T$$=$$0$ axis at $p$$=$$p^{\ast}$, 
depicts the first-order phase transition. In the overdoped region, $p/p_{\rm cr}>1$, the boundary of the superconducting dome marks the region where the magnetic monopoles cease to exist. Above this phase boundary, the Cooper pairs are free to split into single electrons to form a normal metal like in an ordinary superconductor. In the underdoped region $p/p_{\rm cr}<1$, the phase boundary marks the temperature at which the fundamental charge condensate vanishes, but the charged monopole condensate survives at higher temperatures. This implies that the oblique superinsulator forms above the dome. We identify this state as the pseudogap state that sets in between the dome and the temperature $T^{*}$\,\cite{Barisic2013}. 
Note that the charge condensate disappears at $T>T_{\mathrm c}$ not because of the collapse of the coexisting monopole-Cooper pair mixture to the true ground state, as one would have expected in the case of the adiabatic over-cooling transition, but because Cooper pairs excitations that now exist in an out-of-condensate state acquire a large mass. In other words, Cooper pairs viewed as charge excitations acquire an energy gap in their excitation spectrum above $T_{\mathrm c}$, hence do not contribute to transport in precise analogy to usual electrons in a usual insulator. In the proposed model, the physics of the high-$T_c$ cuprates is dominated by the {\it tricritical point} at $p/p_{\rm c}=1$ and $k_{\rs B}T_{\rm c}/\Lambda = 4.9$, where the first-order transition within the superconducting dome meets two continuous transitions to a normal metal (in the overdoped region) and to the pseudogap state (in the underdoped region). This universal structure is an implication of the presence of the magnetic monopole condensate in cuprates. 
Furthermore, monopoles ensure that in the underdoped region the charge carriers appear as paired well above $T_c$\,\cite{Bozovic2019}: the out of condensate Cooper pairs are protected from splitting into single electrons by the Dirac quantization condition. 
\bigskip

The key ingredients of HTS and its elevated transition temperature emerging from our model are magnetic monopoles and the magnetoelectric effect encoded in the $\theta$-term. Both of them are of the topological origin and both are prominently 3D effects. A word of caution is, however, needed at this point. When discussing dimensionality effects in HTS one must be very careful to specify precisely which effect one is addressing. It is well established, e.g. that the electric response in the fluctuation region is described by the 2D behavior of the resistivity and the Hall coefficient, see\,\cite{Pomar1996,Zhao2019} and references therein, reflecting the dominant quasi-2D nature of the macroscopic electric response of cuprates. At the same time, the topological effects are of the 3D nature. This is perfectly consistent with the fact that the unit cell of the HTS materials contains two neighboring CuO planes, which makes it possible for the Cooper pairs not only to move along the $ab$-planes but also to tunnel between them in the $c$-direction. Therefore the HTS materials, even monolayers containing a single unit cell, are essentially 3D with respect to monopole formation and host real 3D quantized Dirac monopoles. These monopoles carry a half of the fundamental magnetic charge since the corresponding unit charge is the charge of the Cooper pairs. A single electron, in the presence of these half monopoles, would have acquired an additional spin $S= \hbar (eg/4\pi) = \hbar/4$, which is impossible due to the non-Abelian nature of the rotation group in 3D. As a result, the emergence of single electrons is prevented by the energy barrier associated with this topological restriction and is suppressed in the superconducting phase of HTS materials. This topological obstruction ties $T_{\mathrm c}$ self-consistently to the condensation temperature of magnetic monopoles, independently of the pairing gap.  At this point, we can conjecture that this superconducting transition temperature $T_{\mathrm c}$ should coincide with the temperature of the formation of the monopole condensate, $T_{\rs M}$.
However, a precise self-consistent derivation of the enhanced transition temperature $T_{\mathrm c}$ requires a microscopic theory of Cooper pairing in the presence of half-magnetic monopoles, which goes beyond the scope of this work and is a subject of a forthcoming publication. Note finally that, at variance to what occurs in 3D HTS systems, magnetic monopoles in 2D are not real particles but rather tunneling events, instantons, which do not impose any topological obstruction on Cooper pair breaking. 
This explains why even a monolayer of HTS material has essentially the same transition temperature $T_{\mathrm c}$ as the bulk sample\,\cite{Yu2019} in stark contrast to behaviors of conventional 2D superconducting films which do not harbor real 3D monopoles. As a result, conventional superconducting films, even hosting the SIT and the superinsulating state, demonstrate a significantly reduced $T_{\mathrm c}$ compared to that of the bulk samples of the same materials.

\section{The nature of the pseudogap state}~~
In a superconductor, the electromagnetic response is mediated by photons with mass $(1/2) \lambda_{\rs L}^{-2} A_{\mu} A^{\mu}$, leading to the London equations $j^{\mu} = \lambda_{\rs L}^{-2} A_{\mu}$. The corresponding electromagnetic response of an oblique superinsulator has been derived in\,\cite{qt,dqt}. As in superconductors, the condensate generates a massless mode that combines with the photon to produce a massive excitation. Contrary to superconductors, however, this massless mode is not a scalar representing the phase of the charge order parameter but, rather, the ``dual magnetic phase", embedded into an antisymmetric tensor gauge field $B_{\mu \nu}$, with the three-tensor field strength $H_{\mu \nu \rho} = \partial_{\mu }B_{\nu \rho} +\partial_{\nu} B_{\rho \mu} + \partial_{\rho} B_{\mu \nu}$ invariant under the gauge symmetries of the second kind, $B_{\mu \nu} \to B_{\mu \nu} +\partial_{\mu} \lambda_{\nu} -\partial_{\nu} \lambda_{\mu}$.  Since $B_{0i}$ are Lagrange multipliers with no time derivatives appearing in the kinetic term $H_{\mu \nu \rho}$, and the gauge function $\lambda_{\mu}$ is itself defined only up to a derivative, the antisymmetric tensor contains indeed a single degree of freedom.
The full electromagnetic response of the oblique superinsulator is 
\begin{equation}
	{\cal L} = -{1\over 4f^2} \left( B_{\mu \nu} +F_{\mu \nu} \right) 
	\left( B^{\mu \nu} + F_{\mu \nu} \right)+ \nonumber 
\end{equation}
\begin{equation}
	+ {\theta \over 32\pi^2} \left( B_{\mu \nu} +F_{\mu \nu} \right) \epsilon^{\mu \nu \alpha \beta} \left( B_{\alpha \beta}+F_{\mu \nu} \right)
	+ {1\over 12 \Lambda^2} H_{\mu \nu \alpha}H^{\mu \nu \alpha} \ .
	\label{infmass}
\end{equation}
Since fields enter only in the combination $B_{\mu \nu}$$+$$F_{\mu \nu}$, the gauge invariance of the second kind of $B_{\mu \nu}$ is ensured by the simultaneous shift $A_{\mu }$$\to$$A_{\mu  }$$-$$\lambda_{\mu}$, leading to three degrees of freedom of mass, see Methods,
\begin{equation}
	m_{\theta} = {f \Lambda \over 4\pi} \sqrt{ \left( {4\pi \over f^2} \right) ^2 + \left( {\theta \over \pi} \right) ^2} \,.
	\label{topmass}
\end{equation}
Two of them belong in the original gauge field and one is from the new field due to the monopole condensate. Alternatively, the original electromagnetic field tensor $F_{\mu \nu}$ can be reabsorbed by the new field in a generalized St\"uckelberg mechanism\,\cite{qt} dual to the familiar Anderson-Higgs mechanism. 
The St\"uckelberg mechanism is the Anderson-Higgs mechanism without a Higgs field or, in other words, the limit of the Anderson-Higgs mechanism in which the Higgs field becomes infinitely heavy and predates both the Anderson and the Higgs versions. That the phase of the order parameter and the gauge field appear  as a combination $(\partial_{\mu}\varphi -A_{\mu})$, means that it is the St\"uckelberg mechanism, much anterior to Anderson’s realization, that works. Usually, it is the photon that ``eats up” the phase of the order parameter. The dual St\"uckelberg formulation, in which the scalar is encoded in the antisymmetric tensor field, presents an opposite situation. 
To conclude here, the electromagnetic response of the oblique superinsulator expresses fully in terms of the antisymmetric tensor field with three degrees of freedom and mass given by Eq.\,(\ref{topmass}).
\bigskip

The mass (\ref{topmass}) is the sum of normal and topological contributions. In the limit $f$$\gg$$1$, the topological contribution dominates and the mass eventually diverges as $f\to \infty$, $\Lambda \to \infty$.  In this limit, the bulk dynamics is frozen because only the topological contribution, 
\begin{equation}
	{\cal L}_{\rm top} = {\theta \over 32\pi^2} \left( B_{\mu \nu} +F_{\mu \nu} \right) \epsilon^{\mu \nu \alpha \beta} \left( B_{\alpha \beta}+F_{\mu \nu} \right)  \ ,
	\label{topological}
\end{equation}
survives in the Lagrangian. The topological regime is realized deep in the underdoped region, where $f$$\gg$$1$. 
\bigskip

Let us now consider the oblique superinsulator on an open manifold $M$ with the boundary $\partial M$. The fields $B_{\mu \nu}$ and $A_{\mu}$ in Eq.\,(\ref{infmass}) contain transverse and longitudinal modes, the latter being encoded in the gauge field $\lambda_{\mu}$ defined by $B_{\mu \nu} = \partial_{\mu} \lambda_{\nu} -\partial_{\nu} \lambda_{\mu}$ and the scalar $\xi$ defined by $A_{\mu} = \partial_{\mu} \xi$. In the topological limit the mass (\ref{topmass}) diverges and all bulk transverse modes get frozen. The longitudinal modes, however, contribute a total derivative into Eq.\,(\ref{topological}). When the model is defined on a bounded space, this total derivative for the longitudinal modes is all that remains and it gives rise to a boundary theory governed by the Lagrangian
\begin{eqnarray} 
	{\cal L}_{\partial M} =  {\theta \over 8\pi^2} \lambda_{\mu } \epsilon^{\mu \alpha \nu} \partial_{\alpha} \lambda_{\nu}  +{\theta \over 2\pi} \lambda_{\mu} \Phi^{\mu} -{M\over 2} v^2 \left( \partial_i \xi \right) ^2  - {\theta^2 \over 2\pi^2 M} b^2 
	\label{boun1} \\
	= {\theta \over 8\pi^2} \lambda_{\mu } \epsilon^{\mu \alpha \nu} \partial_{\alpha} \lambda_{\nu} + {\theta \over 2\pi} \lambda_0 \Phi^0 
	-{\theta \over \pi} b \dot \xi 
	-{M\over 2}  v^2 \left( \partial_i \xi \right) ^2  - {\theta^2 \over 2\pi^2 M} b^2 \ ,
	\label{boun2}
\end{eqnarray} 
where $\Phi^{\mu} = (1/2\pi) \epsilon^{\mu \alpha \beta} \partial_{\alpha} \partial_{\beta} \xi$ represents the vortex current and $b=(1/2\pi)  \epsilon^{ij} \partial_i \lambda_j $ plays the role of the ``charge" canonically conjugate to the phase $\xi$.
The kinetic terms for the surface modes are added-in a posteriori and, correspondingly $M$ is a non-universal mass scale and $v$ is the non-universal propagation speed of the surface modes. 
\bigskip

The Hamiltonian of the boundary theory is derived by setting the canonical momenta $\pi_{\xi}$$=$$-(\theta/\pi) b$ and $\pi_{\lambda_2}$$=$$-(\theta/4\pi^2) \lambda_1$ (as usual in pure Chern-Simons theories, one has a choice of deciding which of two components assumes the role of the coordinate and which one is the momentum). Setting the Lagrange multiplier $\lambda_0$$=$$0$ (Weyl gauge) after recording the Gauss law constraint $b$$=$$-\Phi^0$ it implements, gives 
\begin{equation}
	{\cal H}_{\rm \partial M} = {1 \over 2M} \pi_{\xi} ^2 + {M\over 2} v^2 (\partial_i \xi ) ^2  \ ,
	\label{super}
\end{equation}
which describes a massless degree of freedom as can be seen from the Hamiltonian equations of motion $\left( \partial_0^2 -v^2 \sum_i \partial_i^2 \right) \xi = 0$. The gauge field $\lambda_{\mu} $ is not a dynamical field because there are no ``electric fields" appearing in the Lagrangian, and it is expressed entirely in terms of the vortex configuration via the Chern-Simons Gauss law constraint\,\cite{jackiw2}. This constraint, $b$$=$$-\Phi^0$, implies that the momentum $\pi_{\xi}$ conjugate to the phase $\xi$ is not an electric charge, as it would be in superconductors, but the vortex number itself. However, these boundary vortices carry also unit electric charge (2e), as can be seen by including the electromagnetic coupling $(\theta/2\pi) A_{\mu} \Phi^{\mu}$. They are thus dyons themselves. Finally, it follows from the formulation (\ref{boun1}) of the boundary Lagrangian, that these dyons acquire fractional statistics via the boundary Chern-Simons term\,\cite{anyons}. In particular, for the relevant case $\theta =2\pi$ they become fermions. Thus, the topological limit of the oblique superinsulators for $\theta = 2\pi$ contains only boundary states which are massless fermionic dyons carrying unit magnetic and electric charges. The canonical structure and the massless character of these boundary fermions rests on the U(1) combined gauge symmetry $\lambda_{\mu} \to \lambda_{\mu} + \partial_{\mu} \chi$ and $\xi \to \xi-\chi$ inherited from the bulk. Any boundary perturbation that leaves this symmetry intact does not affect the dynamics of the boundary fermions, which are symmetry-protected surface states. Boundary perturbations that break time-reversal symmetry can affect their statistics by effectively changing the value of $\theta$. As a consequence, these surface fermions behave as a Fermi liquid. 

\section{Discussion and Conclusion}~~
The developed theory provides a consistent, unified insight into the observed universal properties of the PG state and adjacent phases emerging near the tricritical point. First, the symmetry-protected surface fermions in the dyon condensate form a Fermi liquid hence should exhibit resistivity $\rho \propto T^2$, which is exactly what is seen in the experiment\,\cite{Barisic2013}.
As derived above, dyons are intimately associated with the presence of the $\theta$-term in the electromagnetic response, presented by Eq.\,(\ref{theta}), which naturally explains the onset of magnetoelectric effects in the pseudogap state, exactly as in strong topological insulators, see\,\cite{Essin2009}. 
The observation that superconducting fluctuations and quantum correction contributions to the resistive behavior near the dome are perfectly described by the standard 2D formulas, see for example,\,\cite{Pomar1996,Zhao2019}, indicates that the parameters of the boundary fermions are close to those of the standard normal quasiparticle excitations in the corresponding materials. The fact that the structurally simple model compound Hg1201 exhibits negligible residual resistivity\,\cite{Barisic2013} complies perfectly well with the notion that for the symmetry-protected 2D boundary fermions localization is absent. Importantly, while the cuprates exhibit quasi-2D physics, the dyon condensate responsible for HTS properties and the phase diagram structure is essentially of three-dimensional origin. This seeming contradiction is immediately resolved by the fact that the two adjacent CuO planes that present the unit cell of cuprates are enough to ensure the existence of Dirac monopoles and the 3D-like interaction between monopoles and Cooper pairs. This explains why even a few- and monolayer cuprates may exhibit practically the same superconducting transition temperature $T_{\mathrm c}$ as bulk samples\,\cite{Zhao2019,Yu2019}. 
\bigskip

Turning to the detailed Hall resistance experimental data\,\cite{bozovicPNAS}, we recall that in optimally doped cuprates the sign reversal of the Hall resistance is attributed to the contribution of vortices carrying in the core excessive electric charge\,\cite{Vinokur_Hall,Zhao2019}. It is natural to conjecture that, in the underdoped regime in the vicinity of the quantum critical point marking the onset of the superconducting dome, dyons that carry both electric and magnetic charge take up on the role of vortices with the excessive electric charge. Moreover, recalling further that the data on the charge Berezinskii-Kosterlitz-Thouless transition into the superinsulating state\,\cite{Mironov2018} indicate that the boundary fermionic states are harbored by the percolation Chalker-Coddington structure, which is known to be a natural host for gapless boundary states\,\cite{Ochiai2015}, one naturally expects erratic behavior with noticeable random components, exactly as observed in the experiment\,\cite{bozovicPNAS}. However, further theoretical study of the dyon-related Hall effect for a quantitative description of the experiment is needed.

\bigskip
An underdoped antiferromagnetic phase of cuprates holding exactly a single hole per Cu site, realizes a 2D spin square Heisenberg antiferromagnet. Thus, the symmetry of the lattice is $C_4$. In such a lattice the adjacent spins are opposite while the diagonal ones are parallel. As a result, the combined electric/magnetic symmetry of the lattice reduces to $C_2$. Since due to the $\theta$-term, the boundary fermions are dyons and carry both electric and magnetic charge, they feel the combined $C_2$ symmetry, hence nematicity with the nematic director going along the CuO lattice diagonals. This essential characteristic of the PG state\,\cite{Kivelson1998,Matsuda2017} goes hand-in-hand with the $T^2$ resistance. 

\bigskip
Now we discuss the linear, $R_{\rs\square}$$\propto$$T$, resistance behavior in the overdoped region, observed at high magnetic fields destroying the Cooper pair condensate at low temperatures and declared a major puzzle of condensed-matter physics\,\cite{Legros2019}. While its generic character in HTS is established experimentally and is associated with a universal scattering rate, its origin remains a mystery. A careful experimental analysis\,\cite{bozovicScience} reveals that the slopes of the linear terms in the low-temperature regime and in the high-temperature, inherently metallic phase, are noticeable different. Moreover, at sufficiently low temperatures and high magnetic field $B$ there is an intercept which is also linear in $B$\,\cite{bozovicScience}, so that the resistivity assumes the form $\rho = a T + b B$. These two can be considered as thermal and quantum contributions, respectively. 

\bigskip
The resolution of the linear-$T$ puzzle lies in the fact that in the presence of magnetic monopoles, the out-of-condensate Cooper pairs behave as fermions\,\cite{ds1, ds2}. At low temperatures, these fermions scatter primarily by bogolons, the coherent fluctuations of the 
residual dyon condensate domains\,\cite{bogolon}. At temperatures higher than the Bloch-Gr\"{u}neisen temperature $T_{\rs BG}\simeq 2sk_{\rs F}/k_{\rs B}$, where $k_{\rs F}$ is the Fermi momentum of the fermionic Cooper pairs and $s$ is the sound velocity of the bogolons, the resistance due to this scattering mechanism is linear in $T$, $R_{\rs\square}$$\propto$$T$\,\cite{bogolon}. In HTS the zero-temperature superfluid density $n_{\mathrm s}$, and, accordingly, the effective `Fermi energy,' $\epsilon^{\ast}_{\rs F}\simeq n_{\mathrm s}\xi/4m^*$, associated with fermionic Cooper pairs are relatively small\,\cite{Emery1995} (here $\xi$ is the superconducting coherence length, and $m^*$ is the effective mass of the carriers). Accordingly, the Bloch-Gr\"{u}neisen temperature $T_{\rs BG}$ is low, especially if the bogolon sound velocity $s$ is also small. Thus, the universal scattering time $\tau\sim\hbar/k_{\rs B}T$ (in physical units) for $T>T_{\rs BG}$ arises similarly to the usual high temperature electron-phonon scattering time. To understand the origin of the different slopes at low and high temperatures, note that the linear dependence can be understood in the framework of the Anderson orthogonality catastrophe (AOC) approach. Indeed, one expects that all the relevant microscopic times are shorter than the scattering time $\tau$. According to the AOC, $\tau$ is related to the heat, $Q$, generated by the scattering processes by the universal relation\,\cite{Lebedev2020} $\hbar/\tau=fQ$, where $f\simeq{\cal O}(1)$ is some numerical function of the microscopic parameters. Adopting $Q=k_{\rs B}T\delta S$, where $\delta S$ is the entropy change associated with the scattering and weakly (logarithmically) depending on temperature\,\cite{Lebedev2020}, one arrives at the universal relation $1/\tau=\alpha k_{\rs B}T/\hbar$, $\alpha={\cal O}(1)$,  giving rise to the linear-in-$T$ resistivity. Since the microscopic origin of scattering at high and low temperatures is different, one expects the different scattering-related entropy changes, hence different slopes.

\bigskip
At the high magnetic field limit, the number of the available degenerate ground states occupied by bosons in the area $A$ perpendicular to the magnetic field $B$ is $N=(1/2\pi)(eB/\hbar)$ (see, e.g.\,\cite{Abrikosov}). This is the number of available scattering centres at $T=0$ and the same reasoning that leads to the linear-in-$T$ resistance leads to the observed linear-in-$B$ behaviour of the intercept. In other words, at high fields the role of the temperature $T$ is taken by $\hbar\omega_{\mathrm c}$, where $\omega_{\mathrm c}=eB/m^*c$ is the cyclotron frequency\,\cite{VinVarl}. 

\bigskip
Finally, as we have mentioned above, the boundary fermionic states are harbored by the percolation Chalker-Coddington structure in accord with experimental findings of\,\cite{bozovicPNAS}. It implies that one can identify $T^*$ with the temperature  at which the "bubbles" hosting massless boundary dyon states form finite clusters but do not extend across the whole system. Then the $T^{**}$ line in the phase diagram is the line of the percolation transition at which the Chalker-Coddington structure spreads across the system and the pseudogap state with nematicity, the $T^2$ resistance and magnetoelectric effects is fully formed.






\section{Appendix}

\subsection{Lattice BF term}
To formulate the gauge-invariant lattice $BF$-term, we follow\,\cite{dst} and introduce the lattice $BF$ operators
\begin{equation}
	K_{\mu \nu \rho} \equiv S_{\mu }\epsilon _{\mu \alpha
		\nu \rho } \Delta _{\alpha } \ ,\,\,\,
	\hat K_{\mu \nu \rho} \equiv \epsilon _{\mu \nu \alpha \rho}
	\hat \Delta_{\alpha }\hat S_{\rho } \ ,
	\label{kop}
\end{equation}
where 
\begin{eqnarray}
	&&\Delta_{\mu } f(x)\equiv {f(x+\ell \hat \mu )-f(x)}
	\ ,\ \  S_{\mu }f(x) \equiv f(x+\ell \hat \mu )\ ,
	\nonumber \\
	&&\hat \Delta_{\mu } f(x) \equiv {f(x)-f(x-\ell \hat \mu )}  \ ,
	\ \ \hat S_{\mu }f(x) \equiv f(x-\ell \hat \mu ) \ ,
	\label{shift}
\end{eqnarray}
are the forward and backward lattice derivative and shift operators, respectively. Summation by parts on the lattice interchanges both the two derivatives (with a minus sign) and the two shift operators; gauge transformations are defined using the forward lattice derivative. The 
two lattice $BF$ operators are interchanged (no minus sign) upon summation by parts on the lattice and are gauge invariant in the sense that
\begin{eqnarray}
	K_{\mu \nu \rho} \Delta_{\nu } = K_{\mu \nu \rho}
	\Delta _{\rho } = \hat \Delta_{\mu } K_{\mu \nu \rho} = 0 \ ,
	\nonumber \\
	\hat K_{\mu \nu \rho }\Delta_{\rho } = \hat \Delta _{\mu }
	\hat K_{\mu \nu \rho} = \hat \Delta_{\nu }
	\hat K_{\mu \nu \rho } = 0 \ .
	\label{gaugeinv}
\end{eqnarray}
They also satisfy the equations
\begin{eqnarray}
	&&\hat K_{\mu \nu \rho} K_{\rho \lambda \omega} =
	-\left( \delta _{\mu \lambda} \delta_{\nu \omega} - \delta _{\mu \omega}
	\delta_{\nu \lambda } \right) \nabla^2  
	\nonumber \\
	&&+ \left( \delta _{\mu \lambda }
	\Delta_{\nu } \hat \Delta_{\omega} - \delta _{\nu \lambda } \Delta_{\mu }
	\hat \Delta_{\omega } \right) 
	+ \left( \delta _{\nu \omega} \Delta_{\mu }
	\hat \Delta_{\lambda } - \delta _{\mu \omega} \Delta_{\nu } \hat
	\Delta_{\lambda } \right) \ ,
	\label{ma} \\
	&&\hat K_{\mu \nu \rho} K_{\rho \nu \omega } = K_{\mu \nu \rho } \hat
	K_{\rho \nu \omega} = 2 \left( \delta _{\mu \omega } \nabla^2 - \Delta_{\mu }
	\hat \Delta_{\omega } \right) \ ,
	\label{maxwell}
\end{eqnarray}
where $\nabla^2 = \hat \Delta_{\mu} \Delta_{\mu}$ is the lattice Laplacian. 
\\
\subsection{Finite Temperature Deconfinement Transition}
In field theory, a finite temperature $T$ is introduced by formulating the action on a Euclidean time of finite length $\beta =1/T$, with periodic boundary conditions (we have reabsorbed the Boltzmann constant into the temperature). If the original field theory model is defined on a Euclidean lattice of spacing $\ell$, then $\beta $ is quantized in integer multiples of $1/\ell$ as $\beta = b/\ell$, with $b$ an integer. 

The lattice Coulomb Green's function $G(0)$ at coinciding points is  
\begin{equation}
	G(0) = {1\over (2\pi)^4} \int_{-\pi}^{\pi} d^4k { 1 \over  \sum_{i=0}^3 4\  {\rm sin} \left( {k^i\over 2} \right)^2 } \ .
	\label{gzero}
\end{equation}	
At finite temperatures, $k^0$ is both defined on a Brillouin zone of length $2\pi $ and also invariant under shifts
$k^0 \to k^0 + 2\pi/\ell \beta$. This can be achieved only by introducing integers $k\in [-b, b]$, called Matsubara frequencies, and restrict the allowed values to $k^0 = \pi k/b$. Correspondingly, integrals over $k^0$ have to be replaced by sums over Matsubara frequencies,
\begin{equation}
	\int_{-\pi  }^{\pi } d k^0 f \left( k^0 \right)  \to \sum_{k=-b}^{k=b} {\pi  \over \beta} f\left( {\pi k\over b} \right) \ .
	\label{summats}
\end{equation}
As a consequence, at finite temperatures $G(0)$ becomes
\begin{equation}
	G(0, T) ={1\over (2\pi)^4}
	\sum_{k=-b}^{k=+b} {\pi \over b} \int_{-\pi}^{\pi} { dk^1 dk^2 dk^3 \over 4\ {\rm sin} \left( {\pi k \over 2b} \right)^2 + \sum_{i=1}^3 4\  {\rm sin} \left( {k^i\over 2} \right)^2 } \ .
	\label{gfinite}
\end{equation}
This affects primarily the parameter $\eta$ which becomes a function of the temperature, $\eta(T) = \pi G(0,T)/\mu$. Introducing
the scale function $S(T) = \sqrt{G(0)/G(0,T)}$ we obtain the result that, at finite temperatures the overall scale of the ellipse (\ref{ellipse}) shrinks by $S(T)$, $\sqrt{1/\eta} \to S(T) \sqrt{1/\eta}$. 
\\

\subsection{Topological two-form mass term in 3D}
Let us consider the following model for an antisymmetric tensor $b_{\mu \nu}$ in 3D\,\cite{qt} 
\begin{equation}
	{\cal L} = {1\over 12 \Lambda^2} h_{\mu \nu \alpha}h^{\mu \nu \alpha} -{1\over 4f^2} b_{\mu \nu}b^{\mu \nu} + {\theta \over 32\pi^2} b_{\mu \nu} \epsilon^{\mu \nu \alpha \beta} b_{\alpha \beta} \ ,
	\label{massterm}
\end{equation}
where $h_{\mu \nu \alpha} = \partial_{\mu} b_{\nu \alpha} + \partial_{\nu} b_{\alpha \mu} + \partial_{\alpha} b_{\mu \nu}$ is the three-form field strength, $\Lambda$ has dimension [mass] and $f$ and $\theta$ are dimensionless. The first two terms are the generalization to two forms of the Proca Lagrangian for a massive vector field. The third term is topological, since it is metric-independent. 

The equations of motions of this model are
\begin{equation}
	\partial_{\mu} h^{\mu \alpha \beta} + {\Lambda^2 \over f^2} b^{\alpha \beta} -{\Lambda^2 \theta \over 8 \pi^2} \epsilon^{\alpha \beta \gamma \delta} b_{\gamma \delta} = 0 \ .
	\label{eqmot}
\end{equation}
Contracting with $\partial_{\alpha} $ we obtain the condition 
\begin{equation}
	\partial_{\mu} b^{\mu \nu} +{f^2 \theta \over 4 \pi^2} h^{\nu} = 0 \ ,
	\label{contr}
\end{equation} 
where $h^{\mu} = (1/6) \epsilon^{\mu \nu \alpha \beta} h_{\nu \alpha \beta}$ is the dual field strength. Finally, contracting (\ref{eqmot}) with $\epsilon_{\nu \gamma \alpha \beta} \partial^{\gamma}$ and using the above condition we get
\begin{equation}
	\left( \partial^2 + m_{\theta}^2 \right) h^{\mu} = 0 \ ,\,\,\,\,
	m_{\theta} = {f \Lambda \over 4\pi} \sqrt{ \left( {4\pi \over f^2} \right) ^2 + \left( {\theta \over \pi} \right) ^2} \ .
	\label{thetamass}
\end{equation} 
This shows that there is a topological contribution to the mass. 

\medskip

\medskip
\textbf{Acknowledgements} \par 
We are delighted to thank Prof.\,Ivan Bo\v{z}ovi\'{c} for illuminating discussions. The work at the University of Twente (V.M.V.) was partly supported by the NWO grant.
M.C.D. thanks CERN, where she completed this work, for kind hospitality.


\vspace{-0.2cm}

\section*{Data availability}
{Data sharing not applicable to this article as no datasets were generated or analyzed during the current study.}



\end{document}